# 改进 YOLOv8n 的高精度轻量化糖尿病视网膜病变检测研究

孙旭菲,　费玉环+,　臧冉,　王庚辰,　苏猛,　刘丰豪

曲阜师范大学 工学院, 山东 日照　276826
+ 通信作者 E-mail: yuhuanfei@qfnu.edu.cn

**摘　要:** 糖尿病视网膜病变的早期检测与诊断是当前眼部疾病研究的重点之一, 然而, 由于微小病灶特征不明显且易受背景干扰, 现有检测方法在精度和鲁棒性方面仍面临诸多挑战。为了解决这些问题, 提出了一种基于改进 YOLOv8n 的轻量化高精度检测模型 YOLO-KFG。首先, 设计新的动态卷积 KWConv 和 C2f-KW 模块改进主干网络, 提升模型对微小病灶的感知能力。其次, 设计特征聚焦扩散金字塔网络 FDPN, 实现多尺度上下文信息的充分融合。提升模型对微小病灶的感知能力。最后, 设计轻量级共享检测头 GSDHead, 减少了模型的参数量, 使得模型更能部署在资源受限的设备上。实验结果表明, 与基础模型 YOLOv8n 相比, 改进后的模型参数量减少 20.7%, mAP@0.5 提高 4.1%, 召回率提升 7.9%。与单阶段主流算法 YOLOv5n、YOLOv10n 等相比, YOLO-KFG 在检测精度与效率上均表现出显著优势。

**关键词:** 糖尿病视网膜病变; 深度学习; 目标检测; YOLOv8n
文献标志码: A　中图分类号: TP587.1; TP391.4

# Research on Improving the High Precision and Lightweight Diabetic Retinopathy Detection of YOLOv8n

FEI Yuhuan+,　SUN Xufei,　ZANG Ran,　WANG Gengchen,　SU Men,　LIU Fenghao

Qufu Normal University, College of Engineering, Shandong, Rizhao 276826, China

**Abstract:** Early detection and diagnosis of diabetic retinopathy is one of the current research focuses in ophthalmology. However, due to the subtle features of micro-lesions and their susceptibility to background interference, existing detection methods still face many challenges in terms of accuracy and robustness. To address these issues, a lightweight and high-precision detection model based on the improved YOLOv8n, named YOLO-KFG, is proposed. Firstly, a new dynamic convolution KWConv and C2f-KW module are designed to improve the backbone network, enhancing the model's ability to perceive micro-lesions. Secondly, a feature-focused diffusion pyramid network FDPN is designed to fully integrate multi-scale context information, further improving the model's ability to perceive micro-lesions. Finally, a lightweight shared detection head GSDHead is designed to reduce the model's parameter count, making it more deployable on resource-constrained devices. Experimental results show that compared with the base

**作者简介:** 孙旭菲(1998一), 女, 硕士研究生, 研究方向为目标检测; 费玉环(1982一), 女, 博士, 副教授, 研究方向为计算机模拟、机器学习; 臧冉(2001一), 女, 硕士研究生, 研究方向为 3D 打印、目标检测; 王庚辰(2001一), 男, 硕士研究生, 研究方向为目标检测; 苏猛(1997一), 男, 硕士研究生, 研究方向为无人机定位、目标检测; 刘丰豪(1999一), 男, 硕士研究生, 研究方向为水下机器人小目标检测。

model YOLOv8n, the improved model reduces the parameter count by 20.7%, increases mAP@0.5 by 4.1%, and improves the recall rate by 7.9%. Compared with single-stage mainstream algorithms such as YOLOv5n and YOLOv10n, YOLO-KFG demonstrates significant advantages in both detection accuracy and efficiency.

**Key words**：diabetic retinopathy; deep learning; object detection; YOLOv8n

糖尿病视网膜病变（Diabetic Retinopathy, DR）是导致视力丧失的主要原因之一，早期会出现微动脉瘤、硬性渗出等微小病灶病变，准确识别这些病灶对疾病的早期诊断至关重要[1,2]。由于 DR 通常呈现尺寸小、对比度低等特征，基于人工特征提取的传统图像分析方法难以实现精准检测[3]。此外，眼底图像的复杂背景、光照变化等进一步增加了检测难度[4]。因此，如何提升小病灶的识别精度、减少漏检率，成为当前 DR 检测研究的关键问题之一[5-7]。

近年来，深度学习在医学检测任务中展现出优异的性能[8]。针对 DR 检测，研究者提出多种基于深度学习的检测方法，主要分为两大类：一类是追求更高检测精度的双阶段目标检测方法，典型代表是 Faster R-CNN[9]和 R-FCN[10]等。例如，Wang[11]等利用 Faster R-CNN 对 DR 进行病灶检测，通过全卷积架构提高计算效率。王嘉良[12]等人设计的基于 R-FCN 自动诊断系统，能够处理任意尺寸的眼底图像，实现糖尿病相关病变的检测。然而，这类方法计算复杂度较高，在资源受限的医疗环境下应用仍受限制。

另一类是追求实时性和计算效率的单阶段目标检测方法，典型代表有 YOLO 系列[13]、Retina Net[14]和 SSD[15]等。例如，Gupta[16]等利用 SSD 结合卷积神经网络构建混合机器学习模型，提出了一种基于智能手机的 DR 检测系统，提高了检测的准确性。Tseng[17]等采用 Retina Net 构建多模态深度学习融合架构进行 DR 检测，增强了模型的泛化能力。Sait[18]等利用 YOLOv7 进行 DR 检测，能够有效减少计算量，提高检测速度和准确性。然而，单阶段方法在微小病灶检测方面仍受特征提取能力限制，导致漏检率较高。

针对双阶段检测计算复杂度高和单阶段检测漏检、识别精度低等问题，本文提出了一种基于改进 YOLOv8n 的 DR 检测方法——YOLO-KFG，主要工作如下：

（1）将 KW(Kernel Warehouse) 卷积引入 YOLOv8n 的骨干网络中，重新构建 KWConv 和 C2f-KW，以增强了模型对微小病灶的感知能力。

（2）设计特征聚焦扩散金字塔网络(Feature-focused Diffusion Pyramid Network, FDPN)，实现多尺度上下文信息的充分融合。

（3）提出轻量级共享检测头 GSDHead，利用 GNConv 结构和共享尺度特征提取权重，有效减少模型参数并提升检测效率。

（4）构建涵盖三类糖尿病视网膜病变的眼底图像数据集，并通过消融实验与对比试验验证 YOLO-KFG

在检测精度、召回率及参数量上的显著优势。

# 1 YOLOv8n 模型

YOLOv8[19]是 Ultralytics 公司在 2023 年发布的 YOLO 系列目标检测模型，在 YOLOv5 的基础上进行了系统性的重构和优化，旨在提升模型的准确性、速度和适应性。YOLOv8 提供了 N、S、M、L、X 五种不同规模的模型变体，以满足不同应用场景对速度和精度的需求。该模型采用统一的模块化架构，由骨干网络（Backbone）、颈部网络（Neck）和头部网络（Head）三部分组成，分别负责特征提取、多尺度特征融合和目标检测的实现。

YOLOv8n 的骨干网络，主要由 Conv、C2f 和 SPPF 三种模块构成。Conv 模块主要用于图像特征的初步提取与下采样处理，其结构由一个 3×3 卷积层、批量归一化以及 SiLU 激活函数构成。C2f 模块通过多层残差路径提取信息并融合不同分支，有助于提升模型对多尺度、多层次特征的表达能力，增强感受野。SPPF 模块采用三次连续的 5×5 最大池化操作，并通过跳跃连接将不同尺度的池化输出拼接后，最终通过 Conv 结构进行整合。该模块的核心优势在于提取图像的多尺度空间信息，增强特征表达的全局感知能力。

YOLOv8n 的颈部网络，采用了融合特征金字塔网络 FPN 与路径聚合网络 PAN 的双向特征融合结构，先通过 FPN 自顶向下采样深层语义特征并与浅层细节拼接，增强低层特征的语义表达，再通过 PAN 自底向上下采样浅层细节并与高层特征融合，强化定位信息。

YOLOv8n 的头部网络，包含三个检测头，每个检测头都包括分类与回归两个独立的分支。该双分支解耦结构有效降低了任务耦合带来的干扰，使分类与检测任务各自独立优化，从而提升了整体检测性能和准确率。

# 2 改进的模型 YOLO-KFG

本文设计的 YOLO-KFG 算法的网络结构如图 1 所示，与 YOLOv8n 相比主要有 3 个改进部分。在骨干网络部分，提出了一种新的 KWConv 卷积和 C2f-KW 算法，用来捕捉不同尺度的特征信息，以聚焦微小病灶特征。在颈部网络部分，借鉴特征聚焦扩散金字塔网络结构的方法，通过扩散机制使特征扩散到各个检测尺度，能让每个尺度的特征都具有详细的上下文信息，更有利于后续目标的检测与分类。在头部网络部分，提出了一种轻量级的共享卷积检测头 GSDHead，通过使用共享卷积，可以大幅度减少参数

量，使得模型更轻量。

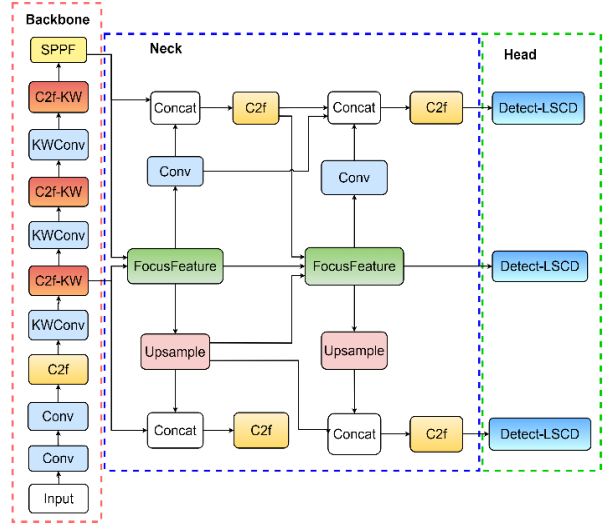

**图 1　YOLO-KFG 结构图**

Fig.1　YOLO-KFG structure diagram

## 2.1　骨干网络改进

在骨干网络中，由于采用固定大小的 Conv 卷积核，难以适应不同尺度的特征，容易造成检测精度低等问题。因此，为了提高微小病灶的检测精度，本文引入一种高效的动态卷积 KW 卷积[20]。KW 卷积的核心理念在于引入了权重生成模块（Weight Generation Module, WGM）和仓库管理器（Kernel Warehouse Manager, KWM）两大核心组件。其中，WGM 通过学习输入特征的分布情况，自适应地生成卷积核的动态权重，使得 KW 卷积具备更强的特征提取能力；KWM 允许多个卷积层共享同一组卷积核，从而在参数管理上更加高效，减少计算资源的浪费，KWM 的表达式如下：

$$Y = \sum_{k=1}^{K} \alpha_k W_k * X_k + b \qquad （1）$$

其中，$W_k$ 表示仓库中的多个共享卷积核，$X_k$ 为输入特征，$b$ 为偏置项，$\alpha_k$ 是由 WGM 生成的动态权重。$\alpha_k$ 使 KW 卷积在不同输入特征下能够自适应调整权重，确保模型在小目标检测中的鲁棒性和精确度。因此，将 KW 卷积引入 YOLOv8n 的骨干网络中，可以大幅提升模型在复杂场景中对微小病灶的检测精度。本文利用 KW 卷积对 Conv 和 C2f 进行改进生成 KWConv 和 C2f-KW，结构如图 2 所示。

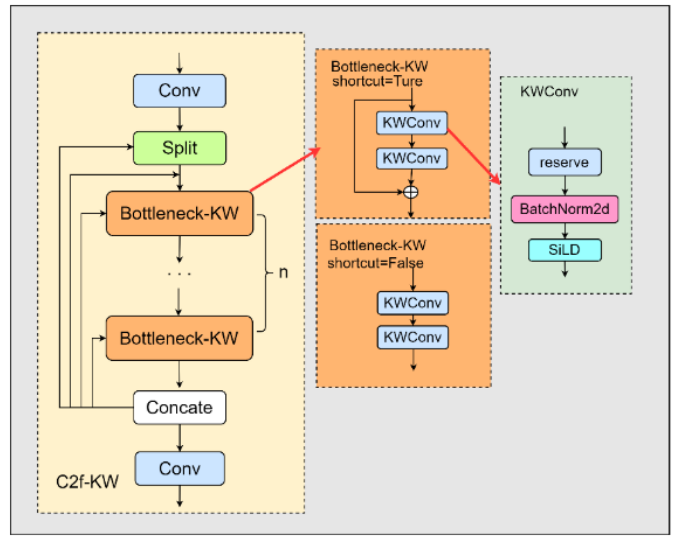

**图 2　KWConv 和 C2f-KW 结构图**

Fig.2　KWConv and C2f-KW structure diagram

首先，将传统的 Conv 卷积与 KW 动态卷积进行有机结合，构建了一种全新的 KWConv 卷积模型。KWConv 卷积在设计上既保留了传统 Conv 卷积的基本流程，又引入了 KWM 这一创新机制，使得多个 KWConv 层可以共享同一组卷积核，从而在参数管理上更加高效，显著减少了参数冗余。同时，KWConv 通过动态调整卷积核权重，使得每个卷积核能够根据输入特征的不同，自适应地调整自身的权重，提高特征提取的灵活性和精确度。

进一步，利用上述思路对 C2f 做了部分改进形成 C2f-KW。最终，将 YOLOv8n 骨干网络中的最后三个 C2f 替换为 C2f-KW 以及最后三个 Conv 卷积替换为 KWConv 卷积。KWConv 卷积和 C2f-KW 模块通过引入 KW 动态卷积，能够提取更加丰富的特征信息，有助于提高微小病灶检测的精度。

## 2.2 颈部网络改进

YOLOv8n 在进行特征融合时，主要依赖 PANet 和 CSP 结构。传统的 PANet 进行信息采集的时候，主要依赖上采样、下采样和横向连接来实现多尺度的特征融合，且得到了最高层特征。层越高，特征图语义信息越丰富，但是分辨率会降低，这样就会不可避免地忽略微小病灶，造成表面特征信息利用效果不佳的问题。为了解决这个问题，本文设计一种全新的特征聚焦扩散金字塔网络结构 FDPN，其结构如图 3 所示。

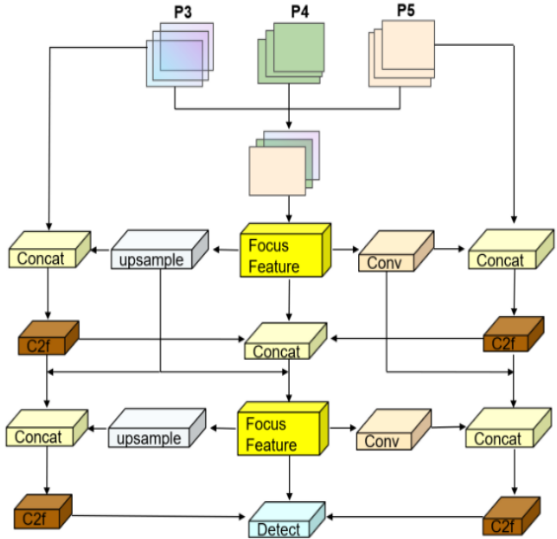

图 3 特征聚焦扩散金字塔网络

Fig.3 Feature focused diffusion pyramid network

FDPN 接收来自不同网络层的三个输入特征，分别通过上采样和 1×1 卷积进行特征对齐后进行拼接，并利用不同尺度的深度可分离卷积提取多尺度信息。随后，通过 1×1 卷积进一步调整特征表示，并采用残差连接以增强梯度流动和信息保留。FDPN 中的特征聚焦模块使得高层次语义信息与低层次细节信息能够相互补充，从而提升模型的检测能力，模块结构如图 4 所示。

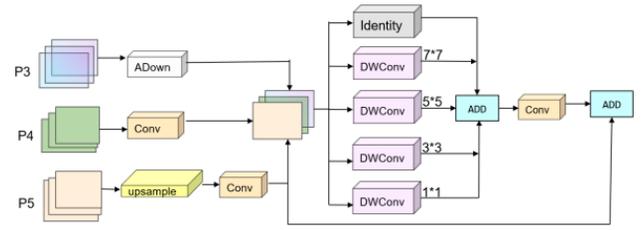

图 4 特征聚焦模块

Fig.4 Feature focusing module

首先，接收来自 P3、P4、P5 这三个层的输入。输入的 P3 层经过 Adown 模块进行特征降采样处理，将输入特征尺寸缩小，同时提取更具代表性的特征，使得低层次的细节信息得到保留。输入的 P4 层则会直接经过一个卷积层，进一步提取中等尺度的语义特征。对于 P5 特征层，先进行一个上采样操作，恢复其空间分辨率，再通过卷积处理，提取高层次的语义信息。其次，把三个尺度处理后的特征信息进行融合，再将融合后的特征信息进行深度可分离卷积（DWConv）和恒等映射（Identity）处理。在此过程中，1×1 的卷积层专注于提取更加细腻和局部的细节特征，而 7×7 的卷积则更侧重于捕获更大范围的上下文信息，从而为模型提供更广泛的理解。最后，将所有卷积操作处理后的特征信息进行加和、卷积等处理。经过上述整个过程，特征信息从低到高的多层级特征能够在不同尺度的检测任务之间有机地扩散和传递，有效增强了模型在上下文信息的表达能力。

## 2.3 头部网络改进

YOLOv8n 模型原本采用解耦合检测头，每个分支包含两个 3×3 的标准卷积层和一个 1×1 的标准卷积层。由于标准卷积的参数量和计算量较大，加剧了模型在资源受限环境中的部署挑战。为解决上述问题，本文设计了一种轻量级共享卷积检测头 GSDHead，核心思想是将原有的两个分支合并为一个，并采用共享参数的方式进行特征提取，从而降低计算量，提高检测效率，GSDHead 结构如图 5 所示。

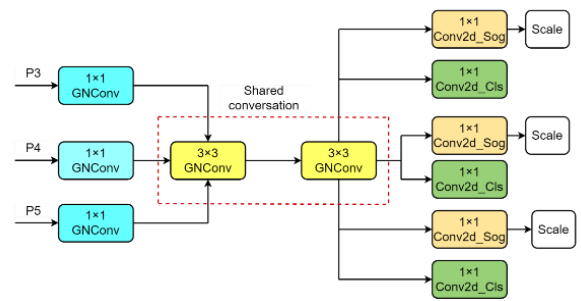

图 5 GSDHead 模型结构

Fig.5 GSDHead model structure

图 5 中的 GNConv[21]是一种全局归一化卷积操作，关键在于在传统卷积的基础上引入全局归一化机制和动态

特征变换，从而提升模型的特征提取能力。本文利用 GNConv 对不同尺度的特征层进行处理，再将特征图输入到由两个 GNConv 组成的共享卷积网络中。GNConv 的引入有助于减少计算复杂度，同时保持较好的特征表达能力。共享卷积通过不同尺度的特征共享信息，减少冗余计算并增强跨尺度特征的交互能力。然后，特征图通过两个不同的卷积层分别计算边界框回归和分类概率，最终通过 Scale 层进行归一化调整，使得梯度更加稳定，提高模型收敛性和数值稳定性。

## 3 实验设计与结果分析
### 3.1 数据集及实验环境

本研究采用自建数据集，数据从 roboflow 网站获取 1325 张具有临床诊断价值的眼底图像，使用 Labelimg 专业工具进行标注。将病变区域划分为"增殖性病变(Proliferate)"、"非增殖性病变(No Proliferate)"及"无病变表征(No-DR)"三大诊断类别。

针对原始图像数量较少的问题，本文对图像进行随机裁剪、伸缩变换、高斯噪声等预处理，将数据集扩增成 6626 张图像，如图 6 所示。

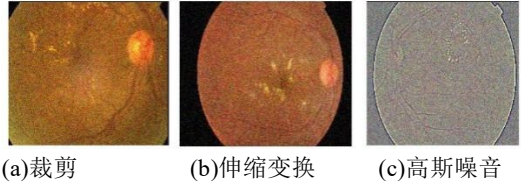

(a)裁剪　　　(b)伸缩变换　　　(c)高斯噪音

**图 6　数据集预处理**

Fig.6　Dataset preprocessing

本实验使用的操作系统是 Windows11，GPU 是 NVIDIA GeForce RTX 4060，编程语言为 Python3.9，深度学习框架为 Pytorch1.10.1，迭代次数为 300 轮。

### 3.2 评价指标

本实验通过精确度（Precision, P）、召回率（Recall, R）、平均精度（mAP@0.5）和每秒帧数（FPS）等指标对目标检测模型进行评估。精度用于衡量预测为正例的样本中，实际正确的比例；平均精度是目标检测中衡量模型综合性能的指标，表示在阈值为 0.5 时所有类别的平均精度均值；而召回率则是以实际样本为依据，表示被正确预测为正例的样本占所有实际正例的比例。检测速度通过 FPS 来衡量，反映模型每秒处理图像的数量。计算精度、召回率和 mAP 的公式如下所示：

$$P = \frac{TP}{TP + FP} \tag{2}$$

$$R = \frac{TP}{TP + FN} \tag{3}$$

$$mAP = \frac{1}{k} \sum_{n=1}^{k} AP_n \tag{4}$$

其中 $TP$ 表示被模型正确识别为阳性的阳性样本数量。$FP$ 表示被模型错误识别为阳性的阴性样本数量。$FN$ 表示模型错误分类为阴性的阳性样本数，mAP 表示类别 n 的平均精度。

### 3.3 消融实验

本文基于 YOLOv8n 模型在数据集上进行了三个核心模块的改进，并针对改进部分进行了消融实验，以验证各组件对模型性能的影响。本文共进行了 10 组实验，具体实验结果如表 1 所示。为确保实验结果的可比性，所有实验在相同的数据参数和环境配置下进行。

**表 1　改进算法在数据集的消融实验**

Table 1　The ablation experiment of the improved algorithm in the dataset

| 实验 | | 精确度(%) | 召回率(%) | mAP@0.5(%) | 参数量(M) |
|---|---|---|---|---|---|
| 1 | YOLOv8n | 85.9 | 84.9 | 90.9 | 3.00 |
| 2 | YOLOv8n+KWConv | 86.0 | 87.3 | 90.1 | 3.02 |
| 3 | YOLOv8n+C2f-KW | 87.6 | 86.5 | 91.2 | 3.04 |
| 4 | YOLOv8n+KWConv+C2f-KW | 90.6 | 89.3 | 94.2 | 3.08 |
| 5 | YOLOv8n+FDPN | 89.3 | 85.8 | 92.1 | 3.04 |
| 6 | YOLOv8n+GSDHead | 89.3 | 86.3 | 92.6 | 2.36 |
| 7 | YOLOv8n+ KWConv+C2f-KW+FDPN | 92.1 | 90.3 | 95.7 | 3.06 |
| 8 | YOLOv8n+KWConv+C2f-KW+GSDHead | 92.4 | 89.2 | 94.6 | 2.44 |
| 9 | YOLOv8n+FDPN+GSDHead | 90.3 | 89.7 | 93.4 | 2.98 |
| 10 | YOLO-KFG | 92.2 | 91.8 | 96.0 | 2.38 |

从表 1 中可以看出，实验 1 即基准模型 YOLOv8n，精确度为 85.9%，召回率为 84.9%，mAP@0.5 为 90.9%，

参数量为 3.00M。单独改进骨干网络的 KWConv 和 C2f-KW，精确度、mAP@0.5 以及参数量与基准模型

几乎持平。当在 YOLOv8n 的特征提取骨干网络最后三个 C2f 替换为 C2f-KW 以及最后三个 Conv 卷积替换为 KWConv 卷积，动态调整卷积核权重，提高特征提取的灵活性和精确度，在数据集上的精确度提高了 4.7%，召回率提高了 4.5%，mAP@0.5 提高了 3.3%。在特征提取网络中引入特征聚集模块，使得高层次语义信息与低层次细节信息能够相互补充，加强微小病灶的特征提取能力，在数据集上的精确度提高了 4.7%，召回率提高了 4.5%，mAP@0.5 提高了 3.3%。引入轻量级共享卷积检测头 GSDHead，将带有全局归一化卷积 GNConv 模块加入到检测头中，实现不同尺度的特征共享信息，减少冗余计算并增强跨尺度特征的交互能力，在数据集上的精确度提高了 3.4%，召回率提高了 1.4%，mAP@0.5 提高了 1.7%，参数量降低了 42.4%。组合实验中，实验 7 的 mAP@0.5 提高了 4.8%，但在参数量方面，与基准模型几乎持平；实验 8 在参数量降低的同时 mAP@0.5 提高了 3.7%，表明动态卷积与轻量化检测头结合可同时提高精确度并降低参数量。改进的 YOLO-KFG 算法模型取得最优性能，与基准模型

型 YOLOv8n 相比，mAP@0.5 提高 5.1%的同时参数量下降 20.7%，且召回率提升 6.9%。

综上所述，在添加动态卷积和 FDPN 模块后，模型的 mAP@0.5 和召回率明显提高，参数量无明显变化，说明这两个模块在提高对微小病灶的特征提取能力的同时不会增加过高的计算复杂度。在轻量化检测头 GSDHead 模块后，观察到模型的参数量明显降低，表明该模块可降低模型的计算复杂度，使得模型更能部署在资源受限的设备上。因此，动态卷积和 FDPN 能有效提升检测精度，轻量化检测头 GSDHead 显著降低了计算成本，三者协同在性能与效率间达到最佳平衡。

## 3.4 对比实验

为了验证改进的 YOLOv8n 算法的检测性能，使用 FPS、mAP@0.5、召回率以及参数量作为定量指标，将其与主流单阶段算法 YOLOv5n、YOLOv8n、YOLOv10n、YOLOv11、YOLOv12n 在数据集上的目标检测结果进行定量分析，结果如表 2 所示。

**表 2 不同网络模型数据集上的性能对比**
Table 1　Performance comparison on different network model datasets

| 方法 | FPS(frame/s) | mAP@0.5(%) | 召回率(%) | 参数量(M) |
|------|------|------|------|------|
| YOLOv5n | 187 | 87.5 | 80.5 | 2.50 |
| YOLOv8n | 171 | 90.9 | 84.9 | 3.00 |
| YOLOv10n | 107 | 86.5 | 79.5 | 2.26 |
| YOLOv11 | 166 | 88.6 | 81.6 | 2.58 |
| YOLOv12n | 173 | 86.8 | 78.5 | 2.50 |
| YOLO-KFG | 187 | 96.0 | 91.8 | 2.38 |

分析表 2 可得，YOLO-KFG FPS 为 187frame/s，与 YOLOv5n 持平，相比于 YOLOv8n、YOLOv10n、YOLOv11、YOLOv12n 分别提升了 8.5%、42.8%、11.2%、7.5%。YOLO-KFG 的 mAP@0.5 值为 96.0%，相比于 YOLOv5n、YOLOv8n、YOLOv10n、YOLOv11、YOLOv12n 分别提升了 8.5%、5.1%、9.5%、7.4%、9.2%。在参数量方面，YOLO-KFG 的参数量仅为 2.38M，相比于 YOLOv5n、YOLOv8n、YOLOv11、YOLOv12n 分别减少了 4.8%、20.7%、7.8%、4.8%。在召回率方面，YOLO-KFG 的召回率为 91.8%，相比于 YOLOv5n、YOLOv8n、YOLOv10n、YOLOv11、YOLOv12n 分别提升了 11.3%、6.9%、12.3%、10.2%、13.3%。

综上所述，YOLO-KFG 在保持较快检测速度的同时，实现了更高的精度和召回率。此外，其参数量更小，模型更为轻量，适用于资源受限的实际应用场景。

## 3.5 可视化分析
### 3.5.1 热力图对比

为了更清晰直观地分析改进模型的检测效果，采用 Grad-CAM++[22]可视化方法分析模型特征图在预测中的作用，通过热力图揭示模型对图像不同区域的关注程度，分析模型的可解释性。热力图中红色越深代表关注度越高，黄色代表关注度次之，蓝色代表对图像检测的影响较小。在图 7(a)中进行检测可以观察到 YOLO-KFG 比 YOLOv8n 具有更好的定位精度，检测目标处呈现更亮的颜色，这说明改进的方法能够更好的聚焦于微小病灶。从图 7(b)中可以看出，YOLOv8n 在有噪声的环境下检测时存在误检的情况，YOLO-KFG 能够有效的检测到被误检的微小病灶且检测检测目标处红色更集中，更易检测微小病灶。从图 7(c)中可以看出，YOLOv8n 在有遮挡的环境下检测时存在漏检的情况，这表明复杂背景对检测效果会产生一定的影响。然而，通过改进的模型 YOLO-KFG 能够有效的检测到被漏检的微小病灶，验证了改进模型能够有效的缓解漏检问题。

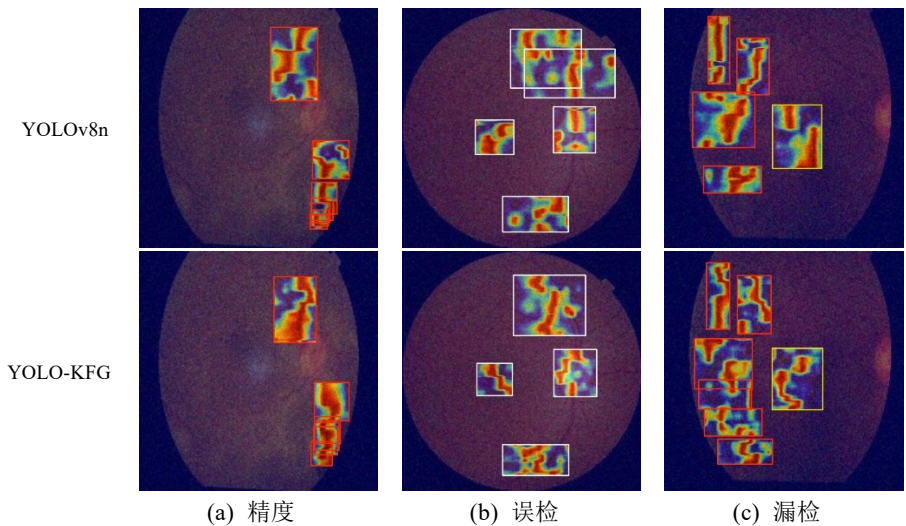

(a) 精度    (b) 误检    (c) 漏检

**图 7    部分测试集改进前后模型热力图**

Fig.7    The model heat map of part test set before and afterimprovement

### 3.5.2    检测结果对比

为了直观展示 YOLO-KFG 算法的检测性能,从测试集中挑选了具有严重遮挡和颜色失真的图像,并将这些图像的检测结果与 YOLOv8n 进行了对比,可视化结果如图 8 所示。

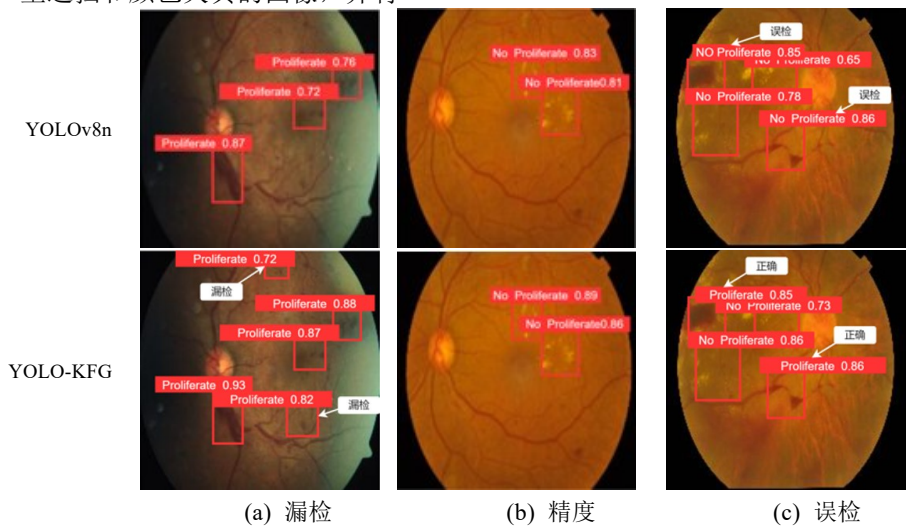

(a) 漏检    (b) 精度    (c) 误检

**图 8    对比实验可视化分析**

Fig.8    Visual analysis of comparative experiments

图 8(a)和(b)显示,YOLOv8n 在检测增殖性病灶时存在两处漏检,而 YOLO-KFG 在相同区域均成功检测(如图 a 中白色箭头所示),表明 YOLO-KFG 在识别病变方面具有更高的精确性。图 8(c)表明,YOLOv8n 还存在两处误检情况,将增殖性病变误分类为非增殖性病变(如图 c 中白色箭头所示),而 YOLO-KFG 能够更精准地识别细微病变区域。此外,YOLO-KFG 的整体检测精度优于 YOLOv8n,其中增殖性病变的检测精度最高提升 15%,非增殖性病变的检测精度最高提升 6%。

综合来看,YOLO-KFG 在 DR 检测上均表现出更高的准确性和稳定性,能够有效降低漏检和误检的风险,尤其在细微病灶识别方面具有明显优势。因此,

在视网膜病变检测任务中,YOLO-KFG 具有更优的临床应用潜力。

## 4    结论

针对 DR 微小病灶检测存在检测精度低、计算效率低等问题,本文提出了一种基于改进 YOLOv8n 的 DR 检测方法。通过引入 KWConv 卷积和 C2f-KW 模块、特征聚焦扩散金字塔网络 FDPN 以及轻量级共享卷积检测头 GSDHead,有效增强了模型对小病灶的识别能力,并降低了计算复杂度。

通过消融实验和对比实验表明,YOLO-KFG 模型相比于原始的 YOLOv8n 模型参数量减少 20.7%,mAP@0.5 提高 4.1%,召回率提升 7.9%。YOLO-KFG

模型在 FPS、mAP@0.5、召回率和参数量方面均优于 YOLOv5n、YOLOv10n 等其他经典单阶段目标检测算法。其中，FPS 最高提升了 42.8%、mAP@0.5 最高提升了 9.5%、召回率最高提升了 13.3%、参数量最高提升了 20.7%。由于 YOLO-KFG 模型在保证高检测精度的同时显著降低了参数量，使其适用于资源受限的边缘设备部署。因此，未来将进一步研究将这一模型部署到边缘设备上，使其能够在实际临床环境中发挥更大的作用，提高远程医疗和基层医疗机构的诊断效率。

## 参考文献：

1. 联系人：费玉环；　2. 通讯地址（邮政编码）：276826；　3. 电子信箱、电话：yuhuanfei@qfnu.edu.cn、18663382322。